\documentclass[pss]{wiley2sp} 
\usepackage{amsmath}
\usepackage{ulem}

\begin{document}


\title{Selfconsistent hybridization expansions for static properties of the Anderson impurity model}

\titlerunning{Ground State of the AIM}

\author{%
  I. J. Hamad  \textsuperscript{\Ast,\textsf{\bfseries 1}}
  P. Roura-Bas \textsuperscript{\Ast,\textsf{\bfseries 2}}
  A. A. Aligia \textsuperscript{\Ast,\textsf{\bfseries 3}}
  E. V. Anda\textsuperscript{\textsf{\bfseries 4}}}

\authorrunning{I. J. Hamad et al.}

\mail{e-mail
  \textsf{roura@tandar.cnea.gov.ar}, Phone:
  +54-11 67727096}

\institute{%
  \textsuperscript{1}\,Consejo Nacional de Investigaciones Cient\'{\i}ficas y T\'ecnicas, CONICET, and Universidad 
  Nacional de Rosario, Argentina.\\
  \textsuperscript{2}\,Depto de F\'{\i}sica CAC-CNEA and Consejo Nacional de Investigaciones Cient\'{\i}ficas
 y T\'ecnicas, CONICET, Argentina.\\
  \textsuperscript{3}\,Comisi\'on Nacional de Energ\'ia At\'omica, Centro At\'omico Bariloche and Instituto Balseiro, 
  8400 S.C Bariloche, Argentina.\\
  \textsuperscript{4}\,Pontificia Universidade Catolica, Rio de Janeiro, Brazil.}

\received{XXXX, revised XXXX, accepted XXXX} 
\published{XXXX} 


\abstract{ By means of a projector-operator formalism we derive an approximation based on 
a self consistent hybridization expansion to study the ground state properties of the 
Anderson Impurity model. We applied the approximation to the general case of finite 
Coulomb repulsion $U$, extending previous work with the same formalism in the infinite-$U$ case. 
The treatment provides a very accurate calculation of the ground state energy and their related zero
temperature properties in the case in which $U$ is large enough, but still finite, as compared 
with the rest of energy scales involved in the model. The results for the valence 
of the impurity are compared with exact results that we obtain from equations derived using the Bethe ansatz 
and with a perturbative approach. 
The magnetization and magnetic susceptibility is also compared with Bethe ansatz results. In order to do this comparison, we also show how to regularize the Bethe ansatz integral equations necessary to calculate the impurity valence, for arbitrary values of the parameters. }

\maketitle   
\section{Introduction}

The Anderson impurity model (AIM) \cite{anderson}, which is one of the most studied Hamiltonians
including strong correlations in condensed-matter physics, has been solved exactly by means of the Bethe ansatz 
\cite{bethe-ansatz,tsvelick}. In addition, the spectral function and many other correlation functions
have been accurately computed by using the numerical renormalization group (NRG) \cite{nrg,nrg-review}.  

However, several approximated schemes are also used. Frequently, they permit 
a rapid scanning of the properties as a function of the model parameters, 
they shed light to the expected behavior and can be used to fit experiments, when due to a large number 
of degrees of freedom it is not possible to use more robust but also time consuming techniques. 
Moreover, there is no Bethe ansatz solution in the case 
of a frequency dependent hybridization function.
Among the approximations, the self consistent hybridization (SCH) expansions for solving the AIM
are commonly used within a large class of different problems. The non-crossing approximation, NCA, \cite{nca} 
which represents the simplest family of these self consistent treatments, provides an accurate calculation
of the Green functions, as well as many other properties, when the Coulomb repulsion is taken
to be infinite. Since the 80's the NCA has been successfully applied to study Ce compounds \cite{nca-Ce}, 
non-equilibrium transport properties \cite{nca-out-eq} and, more recently, non Fermi liquid behaviors \cite{nca-2ch},
electron-phonon interaction  \cite{nca-phonon} and spectroscopy of a double quantum dot system \cite{nca-2QD}.

When the Coulomb repulsion $U$ takes a finite value, the NCA has failed to give the correct Kondo scale ($T_K$).
Unfortunately its value is found to be very underestimated as compared with the correct one
obtained from the Bethe ansatz solution of the model.
The next leading order in the self consistent expansions that partially solves this pathology is often
known as the one crossing approximation, OCA, \cite{oca-1,oca-2,oca-3}. Within this extended formalism 
other classes of
problems have been successfully investigated \cite{oca-usos-impurity}. Among them, its mayor application is in 
the context of the dynamical mean-field theory as an impurity solver \cite{oca-usos-dmft}.

However, the use of NCA or OCA approximations imposes some limitations. Finite temperatures $T$ 
have to be used,
specifically temperatures larger than $0.01 T_K$ in the former and $0.1 T_K$ in the later, in order to 
avoid an artificial increase by abut 10\% in the  spectral weight at the Fermi level.
The same pathology arises when these techniques are applied to systems in which the 
ground state without hybridization is a non degenerate state. This is the case when either the impurity state is 
empty (or formed by an even number of electrons) or when a magnetic field is applied to the impurity 
that brakes the Kramers degeneracy (Zeeman effect).

Remarkably, these limitations when calculating dynamical properties, are absent in the case of static
ones\cite{kang}  like those derived from the ground state energy. In the early approach given by 
Inagaki\cite{inagaki} and Keiter \textit{et al.}\cite{keiter-kimball} the valence of the impurity as
well as the charge and spin susceptibilities were calculated from the ground state energy ($T=0$) 
in the $U$ infinite limit and for a large value of the conduction bandwidth $D\rightarrow\infty$.

Recently, an improvement of these calculations incorporating higher order processes when calculating
the occupancy and magnetization of the impurity was done by three of us\cite{roura-hamad}. The agreement
with exact results was remarkable. 

However, an extension of this static approximation to finite values of $U$ is still desirable. This is
justified due to the existence of relatively not so large values of the Coulomb interaction in 
bulk systems as well as in systems of semiconductor quantum dots or carbon nanotubes
through which transport experiments are currently being performed \cite{edwards,ama,kuba}.

In this contribution we extend the use of the SCH expansions 
when calculating the ground state properties of the single impurity Anderson model to incorporate
large but finite values of the Coulomb repulsion. We compare our approximated ground state properties,
like valence and magnetization of the impurity with exact Bethe-ansatz calculations.
We also provide a method to regularize the divergences that appear in some  Bethe-ansatz expressions 
which might be useful for researchers interested in using these exact results.  
In addition, in the opposite limit, we start the discussion showing the excellent agreement of impurity 
occupancy calculated from perturbation theory (PT) to second order in $U$, modified to 
give the correct result in the atomic limit and to satisfy the Friedel sum rule \cite{kk,none},
for not so large values of $U$ as compared with the hybridization strength.

The paper is organized as follows. In Sec. \ref{modelo}  we introduce the model and,
starting from a projector-operator formalism, 
we review the self consistent hybridization expansion including finite values of the Coulomb repulsion.
we generalize the self consistent hybridization expansion to include finite values of the Coulomb repulsion.
In Sec. \ref{results} we present and discuss the numerical evaluations of different 
physical magnitudes, including a detailed benchmark between Bethe ansatz results and the approximated PT and SCH-expansion. Finally, in Sec. \ref{conclusions} some conclusions are drawn. In the appendix we 
explicitly give the steps that we follow in order to regularize and compute in an efficiently numerical way 
the integral equations of the Bethe ansatz when calculating the impurity valence for arbitrary values of the
model parameters.

\section{Model and formalism}
\label{modelo}

The Anderson Hamiltonian describes a system in which a localized interacting level is coupled with a conduction band of
electrons,

\begin{eqnarray}\label{anderson}\begin{split}
H=& \sum_{k\sigma}\epsilon_{k} n_{k\sigma} + \sum_{\sigma}E_{d} n_{d\sigma} + U
n_{d\uparrow}n_{d\downarrow}+ \\
& \sum_{k\sigma}(V_{k} d^{\dagger}_{\sigma} c_{k\sigma}+ {\rm H.c.}) ,\\
\end{split}\end{eqnarray}
where $n_{k\sigma}=c^{\dagger}_{k\sigma}c_{k\sigma}$ is the number operator and $c^{\dagger}_{k\sigma}$ creates 
a conduction electron with momentum $k$ and spin $\sigma$  in the conduction band. The operators $n_{d\sigma}=d^{\dagger}_{\sigma}d_{\sigma}$ and $d^{\dagger}_{\sigma}$ refer to an electron in the interacting localized impurity state characterized by an energy $E_{d}$ and Coulomb 
repulsion $U$.
The coupling of this level with the bath is taken into account by the hybridization function given in terms of
$V_{k}$ by $\Delta_{}(\omega)\equiv\pi\sum_{k} V_{k}^{2}\delta(\omega -\epsilon_{k})$. 

We will be interested in looking for the energy of the ground state, $E_0$, of the system of $N$-particles satisfying the eigenvalue equation,

\begin{equation}\label{schroedinger}
 H \vert \psi_0 \rangle= E \vert \psi_0 \rangle .
\end{equation}
where $\vert \psi_0 \rangle$ is the ground wave function of the system. 
From it, we can obtain other quantities such as the occupancy of the impurity, magnetization and susceptibility
by differentiation with respect to $E_d$ or to the magnetic field $B$, which can be easily incorporated in this 
formalism via a Zeeman term $g \mu_B B (n_\downarrow-n_\uparrow)/2$ in the Hamiltonian.  

The procedure we follow in general is the same as the one applied in a previous work by three of us \cite{roura-hamad}. 
The main idea is to divide the Hilbert space in two subspaces, by means of projector operators, 
but with the peculiarity that the working subspace consists of a single state. 
For that purpose we define the projector operator $P_1 = \prod_{k\sigma\le k_F}  n_{k\sigma} $, 
where $k_F$ is the Fermi wave vector and $P_2 = 1 - P_1 $. The action of the projectors is then to split the Hilbert space of $N$-particles into two disjoint subspaces, namely $S_1$ which consists of only one state, $\vert \phi_1\rangle =\prod_{k\sigma\le k_F} c^{\dagger}_{k\sigma} \vert 0 \rangle $, the Fermi sea with no electrons in the $d$-level
(which is the ground state in absence of the impurity) 
and $S_2$ with $\vert \phi_2\rangle$ representing the rest of the states in the Hilbert space. 
In principle, the state  $\vert \phi_1\rangle$ can be chosen as the one in which the impurity is empty, singly or doubled occupied, arriving at the same final results.  

The renormalized Hamiltonian that operates on the $S_1$ subspace can be obtained by eliminating 
$\vert \phi_2\rangle $ of the Schr\"odinger equation (\ref{schroedinger}), in a usual procedure \cite{hewson}. 
The action of the effective Hamiltonian on $\vert \phi_1\rangle$ is given by,

\begin{equation}\label{h_eff}
 \tilde{H}\vert \phi_1\rangle = \left(  H_{11} + H_{12}\frac{1}{E-H_{22}}H_{21} \right)
 \vert \phi_1\rangle = E \vert \phi_1\rangle,
\end{equation}
where $H_{ij}=P_iHP_j$. The only contribution to the Hamiltonian $H_{21}$, that connects the subspace $S_1$ 
with $S_2$, comes from the hybridization term in Eq.  (\ref{anderson}). Its operation destroys an electron 
of the Fermi sea and promotes it into the impurity generating a singlet state belonging to the $S_2$ subspace,

\begin{equation}
\vert  \phi_{kd} \rangle =  \sum_{\sigma} \frac{d_{\sigma}^{\dagger} c_{k\sigma}}
{\sqrt{2}}\vert \phi_1\rangle, 
\end{equation}
where the factor $\sqrt{2}$ is included in order to normalize the new state. Here and in what follows,  we employ the notation $k$ and $K$ for wave numbers below and above the Fermi level respectively. We emphasize that the 
state $\vert  \phi_{kd} \rangle$ is, by construction, a singlet 
and therefore its total spin is $S_T=0$. 
Applying $\langle \phi_1 \vert$ to the left of Eq. (\ref{h_eff}) we obtain an equation for the energy $E$,

 \begin{equation}\label{Ene}
 E = \epsilon_T + 2\frac{V^2}{N}\sum_{k, k'} \langle \phi_{kd} \vert \frac{1}{E-H_{22}} 
                     \vert \phi_{k'd} \rangle,                     
\end{equation}
where $\epsilon_T = 2\sum_{k}\epsilon_{k}$ represents the energy of the ground state of the
Fermi sea.  Also, in (\ref{Ene}), we neglect the $k$ dependence of the 
hopping $V_k=V/\sqrt{N}$ as usual, where the scaling by $1/\sqrt{N}$ is necessary to obtain finite 
contributions to the
energy of the Fermi sea-impurity hopping term.

It can be noticed that to get an explicit expression for the eigenvalues in (\ref{Ene}),
it is sufficient to calculate the matrix elements of the operator  $(E-H_{22})^{-1}$ between the states $
\vert\phi_{kd}\rangle$ corresponding to a singlet formed between Fermi sea holes of momentum $k$ and an electron at the
impurity. These states are accessed by only one application of the different $k$ contributions of the Hamiltonian $H_{12}$ to $\vert \phi_1\rangle$ in the $S_1$ subspace.

Furthermore, as we create the space $S_2$ by successive applications of the Hamiltonian to the ground state of the
Fermi sea, since the Hamiltonian commutes with the total spin operator, the space so created contains only states that are singlets.  Although this procedure is not capable of reaching the complete Hilbert space of the system, the ground state, our main object of analysis, is contained in it as it is a singlet.

Up to this point, the procedure has been completely general and equal to the one presented in previous work done 
in the infinite $U$ case \cite{roura-hamad}. Now, in order to generalize the treatment we consider arbitrary 
values of $U$, and hence processes where the impurity is doubly occupied.  
In principle, it is possible to obtain all the matrix elements, $g_{ij}$, of the resolvent operator 
$G = ( E-H_{22} )^{-1}$. However, to obtain the ground state energy, given in (\ref{Ene}),
it is necessary to calculate the diagonal and non diagonal matrix elements $g_{kk'}$ corresponding to the states below the Fermi level. 
The contribution to the energy of the non-diagonal path that starts from a hole in $k$ and finishes with a hole in $k'$ is of the order of $O(V^2/2D^2)$ times less than the diagonal contribution. Following closely the treatment done in \cite{roura-hamad}, and working in the thermodynamic limit, we begin by examining only the diagonal contribution $g_{kk}$, which is expressed 
as a continuous fraction that ultimately, in the thermodynamic limit, can be written in a closed expression as:   

\begin{equation}\label{gkk}
 g_{kk}(E) = \frac{1}{E + \epsilon_{k} - E_d - F_0(E + \epsilon_{k})-F_2(E+\epsilon_k)},
\end{equation}

where the functions  $F_0$ and  $F_2$ satisfy a set of three self consistent equations, exact up to terms proportional to $V^2$.

\begin{eqnarray}\label{f0-f1}
F_0(E) &= \frac{V^2}{N}  \sum_{K} \frac{1}{E-\epsilon_{K}-F_1(E-\epsilon_{K})}\nonumber\\
F_1(E) &= 2\frac{V^2}{N} \sum_{k} \frac{1}{E+\epsilon_{k}-E_d -F_0(E+\epsilon_{k})-F_2(E+\epsilon_k)}. \nonumber\\
F_2(E) &= \frac{V^2}{N} \sum_{k} \frac{1}{E+\epsilon_{k}-2E_d -U -F_1(E+\epsilon_{k})}.
\end{eqnarray}

The system of equations incorporates an extra function,  $F_2$, not present in the infinite $U$ case, related to processes in which the impurity is doubly occupied.

The ground state energy, according to equations (\ref{Ene}) and (\ref{gkk}) and (\ref{f0-f1}), can be self consistently expressed by,

\begin{equation}
 E=F_1(E)
\end{equation}

The self consistent condition represents a non-perturbative treatment to obtain the ground state energy of the 
many-body Anderson Hamiltonian.

Going beyond the diagonal term $g_{kk}$, we incorporate the non diagonal elements $g_{kk'}$,
to the calculation of the ground state energy. Instead of incorporating this crossing terms self consistently into equation \ref{f0-f1}, which is a difficult numerical task, we express this non-diagonal contribution to the energy in terms of the auxiliary functions $F_0$, $F_1$ and $F_2$ calculated as we proposed above.
\begin{equation}
 Q_{1}(E) = 2\frac{V^2}{N}\sum_{k, k'\neq k} g_{kk'}(E),
 \label{correction}
\end{equation}
where the non diagonal element $g_{kk'}$ is  

\begin{eqnarray}\label{non-diag}\begin{split}
 g_{kk'}(E) = \frac{V^{2}}{N} &\frac{1}{E+e_{k}-E_{d}-F_{0}(E+e_{k})-F_{2}(E+e_{k})}\times\nonumber\\
                              &\frac{1}{E+e_{k}+e_{k'}-2E_{d}-U-F_{1}(E+e_{k}-e_{k'})}\times\nonumber\\
                              &\frac{1}{E+e_{k'}-E_{d}-F_{0}(E+e_{k'})-F_{2}(E+e_{k'})}.
\end{split}              
\end{eqnarray}
which gives in total, taking into account the $V^2$ in Eq.  (\ref{correction}), 
a contribution proportional to $V^4$ to the energy. 


When $F_0$, $F_1$ and $F_2$ are calculated through an iterative numerical convergent process imposed by 
the fulfillment of Eq.  (\ref{f0-f1}), it is possible to obtain the ground state energy of the system 
by the solution of Eq.  (\ref{Ene}) that can be written as,

\begin{equation}\label{Ene-final}
 E = F_1(E) + Q_{1}(E)
\end{equation}

In what follows we discuss the impurity properties derived by the numerical solution of the 
Eq.  (\ref{Ene-final}).

\section{Numerical Results}\label{results}

To solve the self-consistent equations, we consider the hybridization function to be a step function 
$\Delta(\epsilon)=\Delta \theta(D-\vert\epsilon\vert)$ where $\Delta=\pi V^2 / 2D$, and $2D$ is the 
bandwidth.
From now on, we set the hybridization $\Delta=1$ as the unit of energy.\\

\textbf{Ground state energy and impurity valence}
\vspace{0.1cm}

As we state in the introduction, it is important for many applications to get a simple approximation 
that can be used, when the Coulomb repulsion is large enough but still finite, for an examination of the 
magnitudes derived from the ground state energy, such as the valence of the impurity. 

\begin{figure}[h!]
\includegraphics[clip,width=7.0cm]{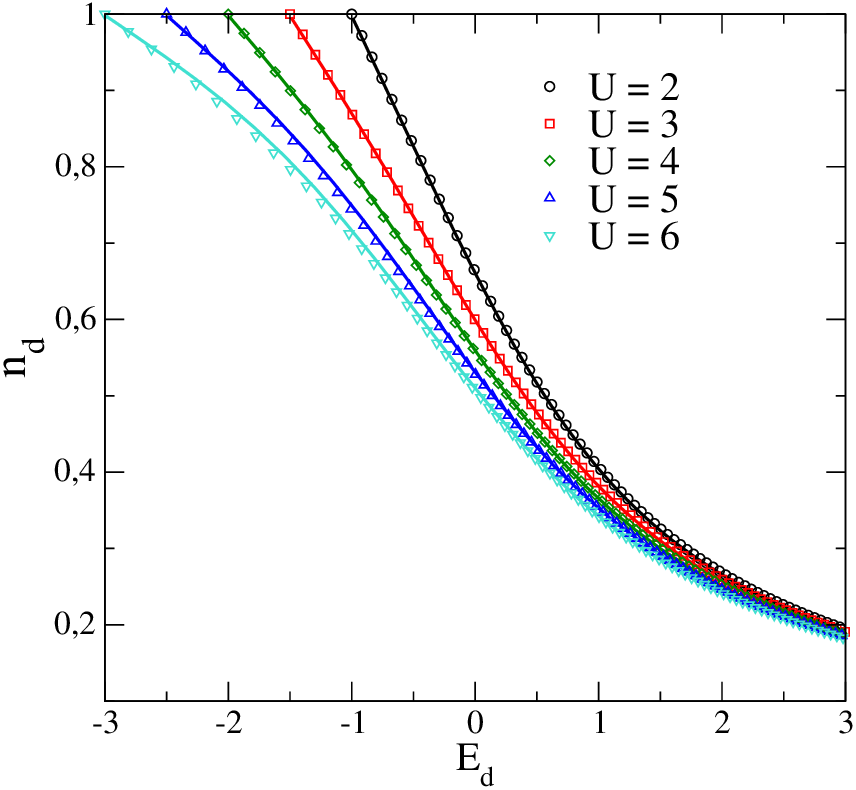}
\caption{(Color online) Total occupation of the impurity as a function of the energy level $E_d$. Solid lines 
stand for the Bethe ansatz results while symbols indicate the results from PT.}
\label{perturbaciones-vs-BA.eps}
\end{figure}

On the other hand, when the Coulomb repulsion is not too large, that is, of the order of the hybridization
energy $\Delta$, perturbation theory in the parameter $U$ provides accurate results. 
For instance, renormalized-PT has been successfully used to analyze the temperature and voltage universal dependence
of the conductance \cite{sca,ng}. 
Specifically, in Fig. \ref{perturbaciones-vs-BA.eps} 
we show the total occupancy of the impurity obtained from PT 
improved in such a way that it reproduces the atomic limit and satisfies the Friedel sum rule 
for all occupancies \cite{kk,none}, and the exact one calculated from the
Bethe ansatz (see appendix) for several values of $U$ as a function of the impurity energy $E_d$.

As it can be seen from Fig. (\ref{perturbaciones-vs-BA.eps}), the impurity occupation obtained 
from PT only presents small deviations as compared with the exact ones (less than $1\%$) when
the Coulomb repulsion is set to be $U=6\Delta$.
Unfortunately, the deviations from the exact values 
quickly grow in the strong interacting limit, when the Coulomb repulsion 
is several orders of magnitude larger than $\Delta$. 
Also our results (not shown) indicate that the magnetic susceptibility within 
the modified PT approach underestimates the magnetic susceptibility by more than $6 \%$ 
for $U \ge 3\Delta$ in the symmetric case $E_d=-U/2$.

This situation emphasizes the convenience of having an approach capable of studying the strong interacting limit, $U>>\Delta$ since it represents a realistic case of most semiconductor and molecular quantum dots. The SCH expansion we propose to calculate the ground state energy, impurity occupation, magnetization and susceptibility satisfies this condition. In what follows we present our main results.

\begin{figure}[tbp]
\includegraphics[clip, width=7cm]{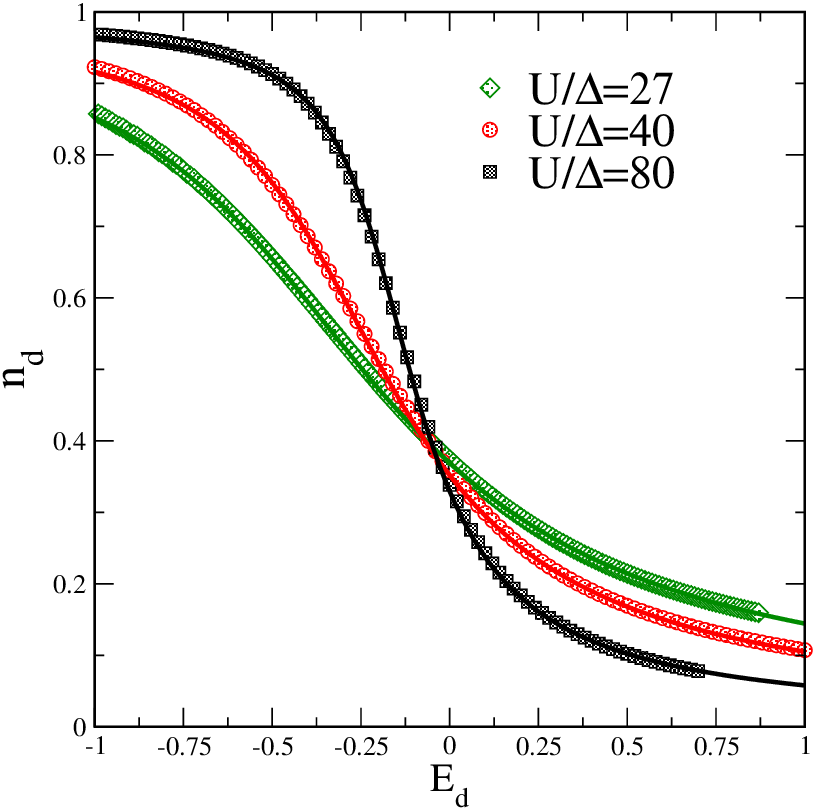}
\caption{(Color online) Total impurity occupancy as a function of the energy level $E_d$
in the strong interacting limit for several values of the Coulomb repulsion $U$.
Solid lines stand for the Bethe ansatz results.}
\label{occ_asymmetric_limit_U8.eps}
\end{figure}

In Fig. (\ref{occ_asymmetric_limit_U8.eps}) 
we show the impurity occupation as a function
of the local level $E_d$ for several values of $U$ 
calculated from the SCH expansion including the 
4th order processes. The solid lines indicate the 
results obtained from the Bethe ansatz technique. We include an appendix in which we explicitly show
the steps we follow to compute the integral expressions that determine the valence within the 
Bethe ansatz solution for arbitrary values of the model parameters. Here, we 
have changed the unit of energy $\Delta=1$ by setting $\Delta=0.1; 0.2$ and $0.3$ while we chose $U=8$.

\begin{figure}[tbp]
\includegraphics[clip, width=7cm]{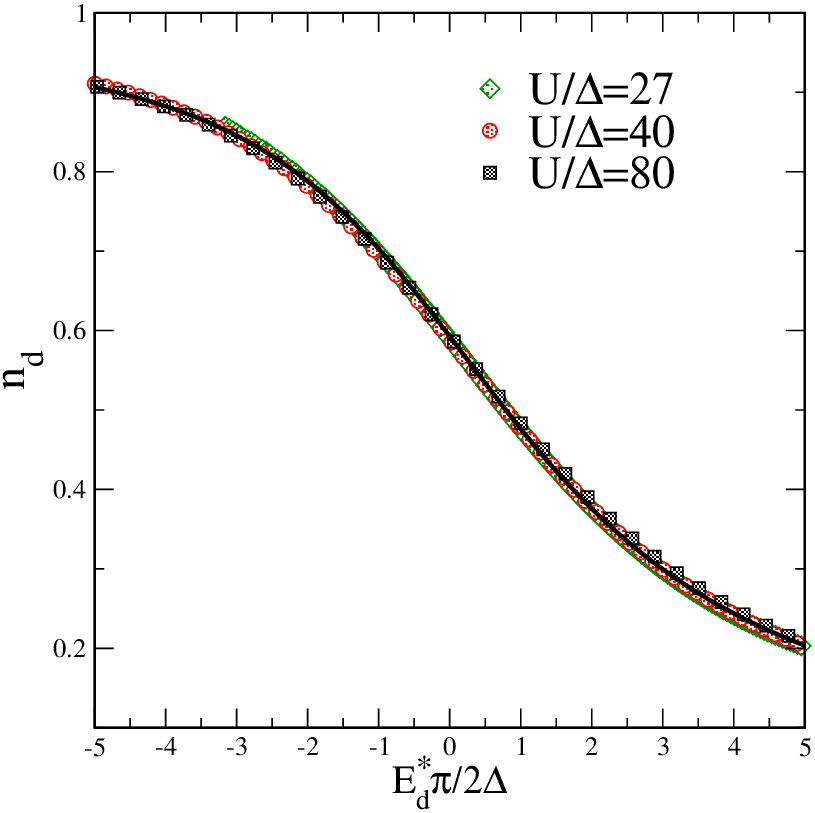}
\caption{(Color online) Universal occupation number $n_d$ as a function of the scaled 
energy $\pi E_d^\ast/2\Delta$ for the same parameters that in Fig. (\ref{occ_asymmetric_limit_U8.eps}).}
\label{Universal_occ_asymmetric_limit_U8.eps}
\end{figure}

In Fig. (\ref{Universal_occ_asymmetric_limit_U8.eps}) we plot the SCH results for the impurity 
occupancy shown in 
Fig. (\ref{occ_asymmetric_limit_U8.eps}) as a function of the renormalized energy 
$E_d^\ast=E_d + \frac{\Delta}{\pi}$ln$(\frac{\pi{e}U}{4\Delta})$ which explicitly
includes the Haldane shift \cite{haldane} together with the corresponding renormalized
Bethe ansatz curve. As expected from the universality of the model, we notice here that all the 
curves collapse into an universal one 
in the Kondo regime $-E_d^\ast \gg \Delta$ 
and that at $E_d^\ast = 0$ the valence is $n_d$ = 0.58 
in agreement with Fig. 2 of Ref. \cite{bethe-ansatz}.\\

\textbf{Magnetic properties}
\vspace{0.1cm}

In what follows, we add a magnetic field $B$ to the system and analyze the 
magnetization of the impurity, $M_{imp}$.
The inclusion of the Zeeman interaction within the SCH procedure is straightforward and it is described in Ref. \cite{roura-hamad}. 
According to the usual definition, \cite{nrg-review}, to obtain $M_{imp}$ we subtract from the total magnetization 
of the system the magnetization of the isolated Fermi sea. 
Therefore, the impurity contribution to the magnetization, $M_{imp}$, is
given by,

\begin{equation}\label{magnetization}
 M_{imp}(B) = - \frac{\partial }{\partial B} (E(B)-\epsilon_T(B)).
\end{equation}

\begin{figure}[tbp]
\includegraphics[clip, width=7cm]{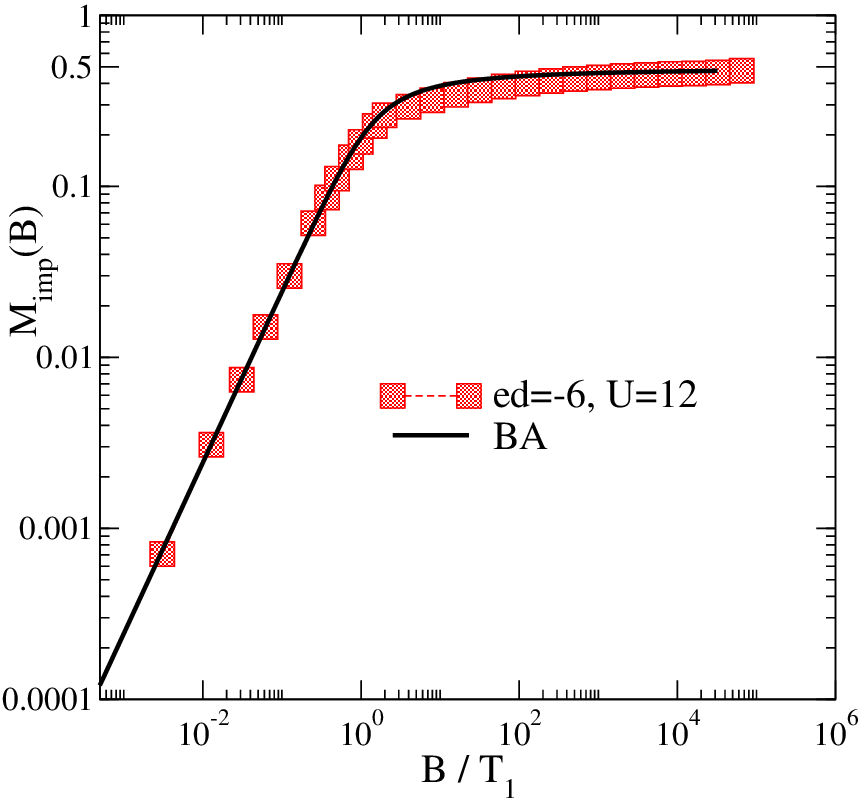}
\caption{(Color online)  Comparison of the results obtained by the SCH expansion for the symmetric case and the Bethe ansatz 
solution of the Kondo model, for the  impurity 
contribution to the magnetization $M_{imp}$ in units of 
$g\mu_B$ as a function of $B/T_1$. The energy impurity level is taken to be $Ed = -6$. The scale $T_1$ is defined in 
the main text. }
\label{magne-BA-2015.eps}
\end{figure}

In Fig. (\ref{magne-BA-2015.eps}), we show the impurity contribution to the magnetization in units 
of $g\mu_B$, as a function of $B$. We chose $E_d=-6$ to compare our results with the exact Bethe ansatz  
results for the Kondo model 
\cite{nandrei}. The selected values of the energy level $E_d=-6$ and $U=12$ in units of $\Delta$, 
corresponds to the symmetric case, in which the charge 
fluctuations are frozen and the Anderson model maps onto the Kondo one.

Following the notation introduced by N. Andrei in Ref. \cite{nandrei}, 
we define the energy scale $T_1$, that
characterizes the strong coupling regime (here it means Kondo regime)\cite{hewson},
from the susceptibility at zero temperature, 

\begin{equation}\label{T_1}
 \chi_{imp}(B=0)\equiv\frac{\sqrt{ 2 \pi e} }{T_1}.
\end{equation}

It can be noticed from Fig. (\ref{magne-BA-2015.eps}) that the agreement between SCH expansion and BA 
is very good when $M_{imp}$ is calculated in units of $B/T_1$, for which $M_{imp}$ is an universal
curve.  
For small values of $B$ as compared with $T_1$, we obtain a linear 
dependence of the magnetization with respect to the applied field. This agrees with the 
expected behavior of the Kondo problem, which is indicative of a screened
impurity spin.  On the other hand, for larger fields, we obtain the characteristic 
slow asymptotic approach to the value $g\mu_{B}/2$ of saturation, which also agrees 
with the well known logarithmic corrections \cite{nandrei,tsvelick,hock}.  

From the definition of the low energy scale $T_1$, it is natural \cite{hewson}
to relate it with the usual Kondo scale, $T_K$ via $T_1 =  \sqrt{ 8 \pi e}~ T_K$. 
Unfortunately, the Kondo scale that we obtain within the SCH procedure is found to be
several orders of magnitude smaller than the expected one from the well known Haldane scale,
$T_K=\sqrt{\frac{U\Delta}{2}}exp(-\frac{\pi U}{8\Delta})$ \cite{haldane}, for the symmetric case.
For the parameters used in Fig. \ref{magne-BA-2015.eps}, the SCH expansion predicts a Kondo 
scale of the order of $10^{-5}$ while the corresponding one to the Haldane expression is
of the order of $10^{-2}$ in units of $\Delta$.

This is not a surprising result. The same feature appears when the SCH expansion in its 
lowest version, NCA, is applied using finite values of the Coulomb interaction 
\cite{oca-1,oca-2,oca-3}. As we mentioned in the introduction, 
this pathology was partially overcome, adding in a self consistent way, all 
the dressed diagrams that involve one crossing conduction electrons, which corresponds 
to the next leading order called one-crossing approximation OCA. While in our present 
treatment of the problem we are already taking into account $4$th order processes, we are
not including these contributions in a self consistent way, as we mentioned in section \ref{modelo} and can be seen from Eq. (\ref{f0-f1})). 
While our treatment is completely reliable to obtain the static properties of the Anderson Hamiltonian in the strong coupling limit, 
it fails to give correctly the Kondo temperature when $U/\Delta$ is not large enough, situation in which an equivalent scheme to OCA approximation is required. This full treatment is beyond the scope of the present study. 

\begin{figure}[tbp]
\includegraphics[clip, width=7cm]{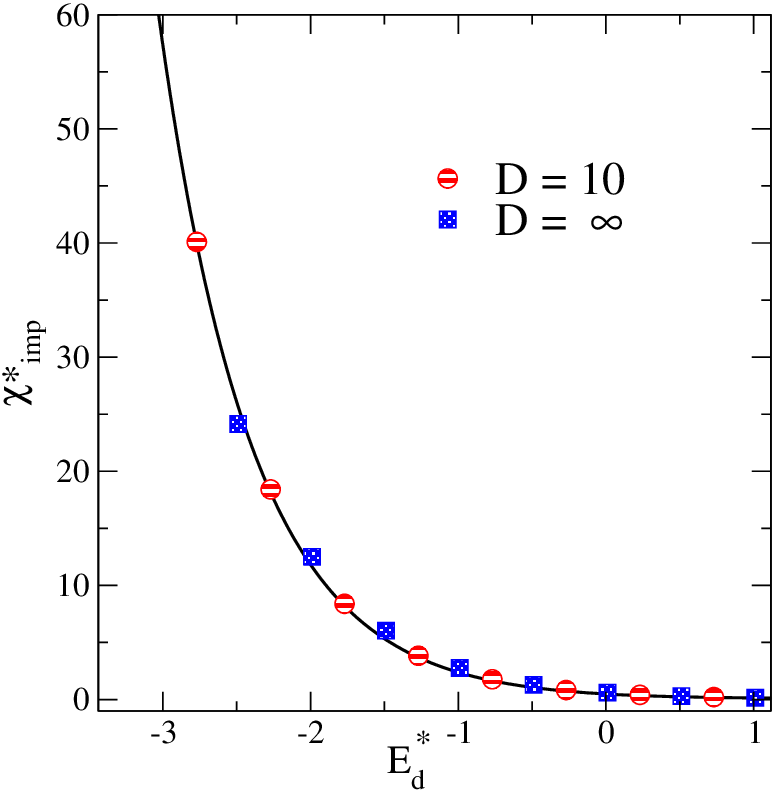}
\caption{Impurity contribution to magnetic susceptibility $\chi^*=\chi/((\mu_B g)^2/\Delta)$as a function of the renormalized energy 
$E_d^\ast$.  The solid line indicates the calculation by using the Bethe ansatz technique in the limit
of both, $D,U\rightarrow\infty$. The symbols represent the SCH solution for two selected values of the 
bandwidth $D$. }
\label{chi_BA_vs_EA.eps}
\end{figure}

As a final analysis, we include in Fig. \ref{chi_BA_vs_EA.eps}
the susceptibility comparing our results for $\chi_{imp}$ with those obtained from a Bethe ansatz calculation \cite{aligia} in the case of infinite Coulomb repulsion. In this case, we show results for both $D=10$ and $D\rightarrow\infty$ properly scaled by a Haldane shift. 
The agreement with the Bethe ansatz solution of $\chi_{imp}$ is very good in both cases, capturing the universality of the model.

\section{Summary and Conclusions} \label{conclusions}

In this work we applied the self consistent hybridization expansions including 4th order processes to compute
the ground state energy, occupancy and magnetic properties of the Anderson impurity model in the case of finite
Coulomb repulsion, extending a previous treatment for infinite $U$. We show that the static properties of the 
model can be well calculated within the SCH expansions for zero temperature and finite magnetic fields. We
successfully match our results with those coming from the Bethe ansatz solution of the model. 

This simple approach allows the study of systems that can be described by the Anderson model, such as  semiconductor \cite{ama} or molecular
quantum dots \cite{kuba} where the Coulomb repulsion is expected to be large but cannot be considered infinite. Experimentally relevant quantities, 
such as the equilibrium conductance of the impurity, which is directly related to the occupancy 
(if the system is in a Fermi liquid regime) can be obtained. Furthermore, the semi analytic nature of the 
solution permits a direct and fast calculation of the properties on a single workstation.

\appendix

\section{Bethe ansatz equations of the impurity valence at zero magnetic field}

To compute the valence of the impurity at zero magnetic field and for arbitrary values
of the parameters $U$, $\epsilon_d$, $\Delta$ we start with the expressions (5.8) and (5.9) given
by Wiegmann and Tsvelick 
in Refs\cite{bethe-ansatz,tsvelick}.

\begin{equation}\label{Z}
\frac{U/2 + \epsilon_d}{(2U\Delta)^{1/2}}=\frac{i}{\sqrt{2\pi}}\int_{-\infty}^{+\infty}
                                          \frac{d\omega}{\omega+i\eta}
                                          \frac{e^{-\vert\omega\vert/2-i\omega Q}}{G^{(-)}(\omega)}
                                          \frac{1}{(-i\omega+\eta)^{1/2}}
\end{equation}
\begin{eqnarray}\label{nd}
n_d = 1 - \frac{i}{\sqrt{2}\pi}\int_{-\infty}^{+\infty}
                                          \frac{d\omega}{\omega+i\eta}
                                          \frac{e^{-\vert\omega\vert/2}}{G^{(-)}(\omega)} \nonumber \\
                                          \times \int_{-\infty}^{+\infty}
                                          d\kappa~e^{i\omega(g(\kappa)-Q)}\Delta(\kappa).                                   
\end{eqnarray}
\\
where the functions $G^{(-)}(\omega)$, $g(\kappa)$ and $\Delta(\kappa)$ are given by

\begin{displaymath}
 G^{(-)}(\omega) = \sqrt{2\pi}~~\frac{(\frac{i\omega+\eta}{2\pi e})^{i\omega/2\pi}}
                   {\Gamma(\frac{1}{2}+\frac{i\omega}{2\pi})},
\end{displaymath}
\begin{displaymath}
g(\kappa) = \frac{(\kappa-\epsilon_d-U/2)^2}{2U\Delta}, 
\end{displaymath}
\begin{displaymath}
\Delta(\kappa) = \frac{\Delta}{\pi}~\frac{1}{(\kappa-\epsilon_d)^2+\Delta^2}, 
\end{displaymath}
with $\eta\rightarrow0^+$.
The equation in (\ref{Z}) fixes the value of $Q$ that enters in the explicit calculation of the 
impurity occupation in (\ref{nd}). 
 
The highly divergent nature of the integrand in Eq.  (\ref{Z}) demands particular care in handling the 
limit $\eta\rightarrow0^+$. 
In fact, using the identity 
$\frac{1}{\sqrt{-i\omega+\eta}}=\frac{1}{\sqrt{-i}}\frac{1}{\sqrt{\omega+i\eta}} 
\rightarrow \frac{1}{\sqrt{2\vert\omega\vert}}~(1+i~sgn(\omega))$ and taking explicitly 
the limit $\eta\rightarrow0^+$ the \small{RHS} of (\ref{Z}), which we call $Z(Q)$, becomes

\begin{equation}
\begin{split}
 Z(Q) =& \frac{i}{\sqrt{2\pi}}\int_{-\infty}^{+\infty}
                                          \frac{d\omega}{\omega+i\eta}
                                          \frac{e^{-\vert\omega\vert/2-i\omega Q}}{G^{(-)}(\omega)}
                                          \frac{1}{(-i\omega+\eta)^{1/2}}\\
      =& \frac{i}{\sqrt{2\pi}}\int_{-\infty}^{+\infty} \frac{d\omega}{\sqrt{2\vert\omega\vert}}~
         (1/\omega-i\pi\delta(\omega)) \nonumber \\
         & \times (1+i~sgn(\omega))~H(\omega,Q)\nonumber,
\end{split}
\end{equation}
with $H(\omega,Q) = \frac{e^{-\vert\omega\vert/2-i\omega Q}}{G^{(-)}(\omega)}\vert_{\eta\rightarrow0^+} = 
\frac{e^{-i\omega Q}\Gamma(\frac{1}{2}+\frac{i\omega}{2\pi})}
{\sqrt{2\pi}\vert\frac{\omega}{2\pi e}\vert^{i\omega/2\pi}~e^{\vert\omega\vert/4}} $ and being the limit
$H(0,Q) = 1/\sqrt{2}$, the argument has a non integrable singularity at $\omega=0$. 
In order to avoid this singularity and make the integral tractable numerically, 
we subtract the singular part of $Z(Q)$ before taking the limit $\eta\rightarrow0^+$.

Let us write the integral $Z(Q)$ in the following form

\begin{equation}
 Z(Q) = \frac{i}{\sqrt{-2i\pi}}\int_{-\infty}^{+\infty}
                                          \frac{d\omega}{(\omega+i\eta)^{3/2}}
                                          \frac{e^{-\vert\omega\vert/2-i\omega Q}}{G^{(-)}(\omega)}.                                       
\end{equation}

Using the fact that  

\begin{equation}
 Z_s =  \frac{ic}{\sqrt{-2i\pi}}\int_{-\infty}^{+\infty}
                                          \frac{d\omega}{(\omega+i\eta)^{3/2}} = 0
\end{equation}
we set $c = 1/G^{(-)}(0)=1/\sqrt{2}$ and re-write $Z(Q)=Z(Q)-Z_s$ as follows

\begin{equation}
\begin{split}
 Z(Q) &= \frac{i}{\sqrt{-2i\pi}}\int_{-\infty}^{+\infty}
                                          \frac{d\omega}{(\omega+i\eta)^{3/2}}
                                          \{~H(\omega,Q)-\frac{1}{\sqrt{2}}~ \}\\
      &= \frac{i}{2\sqrt{\pi}}\int_{-\infty}^{+\infty}
                                          d\omega~
                                          \frac{1+i~sgn(\omega)}{\omega\sqrt{\vert\omega\vert}}~
                                          \{~H(\omega,Q)-\frac{1}{\sqrt{2}}~ \}.\\
\end{split}
\end{equation}

In this way, the argument within the integral becomes regular. 

We also map the infinite range of integration to a finite one by using 

\begin{displaymath}
\begin{split}
\int_{-\infty}^{+\infty}d\omega~f(\omega)&=\int_{0}^{+\infty}d\omega~ \{f(\omega)+f(-\omega)\}\\
               &=
\int_{0}^{1}dx \frac{1+x}{(1-x)^3}\{f(\frac{x}{(1-x)^2})+f(\frac{-x}{(1-x)^2})\}.
\end{split}
\end{displaymath}

The change of variable was chosen so that the argument of the integral is smooth 
for $x \rightarrow 1$ ($\omega \rightarrow \infty$).
By means of the above mentioned transformations, the integral becomes suitable for numerical evaluation.

The integral in (\ref{nd}) is simpler and it can be evaluated in a 
more direct way. Calling $ K(\omega) = \int_{-\infty}^{+\infty} 
d\kappa~e^{i\omega(g(\kappa)-Q)}\Delta(\kappa)$, it becomes

\begin{equation}
\begin{split}
n_d &= 1 - \frac{i}{\sqrt{2}\pi}\int_{-\infty}^{+\infty}
                                          \frac{d\omega}{\omega+i\eta}
                                          \frac{e^{-\vert\omega\vert/2}}{G^{(-)}(\omega)}K(\omega)\\
    &= 1 - \frac{i}{\sqrt{2}\pi}\int_{-\infty}^{+\infty}d\omega~(1/\omega-i\pi\delta(\omega))
                                          \frac{e^{-\vert\omega\vert/2}}{G^{(-)}(\omega)}K(\omega)\\
    &= 1 - \frac{i}{\sqrt{2}\pi}\{ \int_{-\infty}^{+\infty}d\omega~ \frac{e^{-\vert\omega\vert/2}}
           {\omega G^{(-)}(\omega)}K(\omega) - \frac{i\pi}{\sqrt{2}} \}\\
    &= \frac{1}{2} + \frac{1}{\sqrt{2}\pi} \int_{-\infty}^{+\infty}d\omega~\frac{e^{-\vert\omega\vert/2}}
           {\omega}\mathcal{I}m [\frac{K(\omega)}{G^{(-)}(\omega)}]
\end{split} 
\end{equation}

Finally, we compute the principal part of the integral with the help of the following transformations

\begin{displaymath}
\begin{split}
\int_{-\infty}^{+\infty}d\omega~f(\omega)&=\int_{0}^{+\infty}d\omega~ \{f(\omega)+f(-\omega)\}\\
     &=\int_{0}^{1}dx \frac{1}{(1-x)^2}\{f(\frac{x}{1-x})+f(\frac{-x}{1-x})\}.
\end{split}
\end{displaymath}

\begin{acknowledgement}
This work was partially supported by PIP 00273, PIP 01060  and 
PIP 112-201101-00832 of CONICET, and PICT R1776  and PICT 2013-1045 
of the ANPCyT, Argentina. 
We a knowledge financial support from the brazilian agencies CNPq and FAPERJ(CNE). 
\end{acknowledgement}

%
%

\end{document}